\documentclass[twocolumn]{aastex631}

\usepackage{amsmath}
\usepackage{bm}

\usepackage{breqn}

\def\bE{\boldsymbol{E}}

\def\bBbg{\boldsymbol{B}_{\rm bg}}

\def\sigbg{\sigma_{\rm bg}}
\def\bk{\boldsymbol{k}}

\def\Rsh{R_\times}
\def\E{{\cal E}}

\graphicspath{{./}{figures/}}

\begin{document}

\title{%
{Monster Shocks, Gamma-Ray Bursts, and Black Hole Quasi-normal Modes from Neutron-star Collapse}
}

\author[0000-0002-0491-1210]{Elias R. Most}
\affiliation{TAPIR, Mailcode 350-17, California Institute of Technology, Pasadena, CA 91125, USA}
\affiliation{Walter Burke Institute for Theoretical Physics, California Institute of Technology, Pasadena, CA 91125, USA}

\author[0000-0001-5660-3175]{Andrei M. Beloborodov}
\affiliation{Physics Department and Columbia Astrophysics Laboratory, Columbia University, 538 West 120th Street New York, NY 10027,USA}
\affiliation{Max Planck Institute for Astrophysics, Karl-Schwarzschild-Str. 1, 85741 Garching, Germany}

\author[0000-0002-7301-3908]{Bart Ripperda}
\affiliation{Canadian Institute for Theoretical Astrophysics, 60 St. George St, Toronto, Ontario M5S 3H8}
\affiliation{Department of Physics, University of Toronto, 60 St. George St, Toronto, ON M5S 1A7}
\affiliation{David A. Dunlap Department of Astronomy, University of Toronto, 50 St. George St, Toronto, ON M5S 3H4}
\affiliation{Perimeter Institute for Theoretical Physics, 31 Caroline St. North, Waterloo, ON, Canada N2L 2Y5}
\affiliation{Center for Computational Astrophysics, Flatiron Institute, 162 5th Ave, New York, NY, 10010, USA}

\begin{abstract}
\noindent 
We perform the first magnetohydrodynamic simulation tracking the magnetosphere of a collapsing magnetar. The collapse is expected for massive rotating magnetars formed in merger events, and 
may occur many hours after the merger. 
Our simulation suggests
a novel 
mechanism for a 
gamma-ray burst (GRB) 
which is uncollimated and forms
a delayed high-energy counterpart of the merger gravitational waves.
The simulation shows
that 
the collapse launches an outgoing magnetospheric shock, and a hot magnetized outflow forms behind the shock.
The outflow is baryon-free and uncollimated, and its power peaks on a millisecond timescale. Then,
the outflow 
becomes
modulated by the ring-down of the nascent black hole, imprinting its kilohertz quasi-normal modes on the GRB tail.
\end{abstract}


\section{Introduction}\label{sec:intro}

Neutron star mergers are not only sources of gravitational waves, 
but are also accompanied by electromagnetic counterparts. 
Such counterparts provide insights into properties of dense matter in neutron stars 
and their magnetic fields.
Canonical counterparts include a prompt gamma-ray burst (GRB
\citep{Meszaros:2006rc}) and a kilonova
\citep{Li:1998bw,Metzger:2019zeh}. Both were observed from the neutron
star merger GW170817 (see, e.g., \citet{LIGOScientific:2017zic} for a summary). 
The GRB followed the merger with a 2-s delay and was likely emitted by a blast wave breaking out at the photosphere of the merger ejecta \citep{Murguia-Berthier:2017kkn,Gottlieb:2017pju,Xie:2018vya,Beloborodov2020}. 
The kilonova was emitted on a day timescale and powered by nuclear decay in the expanding ejecta. 

The explosion picture usually includes collimated relativistic jets from the black hole promptly formed after the merger.
The jets are extremely bright sources of gamma-rays if observed face-on, and also emit a broad-band afterglow when they decelerate in an ambient medium \citep{Gottlieb:2017pju}.
Similar jets are expected if the merger forms a short-lived  neutron star with a debris accretion disk, see \cite{Rezzolla:2011da,Ruiz:2016rai,Kawamura:2016nmk,Kiuchi:2022nin} for magnetohydrodynamic (MHD) simulations of such systems. Such models predict the collapse of the neutron star in milliseconds to a second after the merger, triggered by accretion or the loss of differential rotation.

However, 
intermediate mass
mergers
may not 
promptly collapse to
black holes and form jets \citep{Bauswein:2013jpa,Koppel:2019pys,Bauswein:2020aag,Kashyap:2021wzs,Tootle:2021umi,Kolsch:2021lub,Schianchi:2024vvi}. Instead, they may form a long-lived neutron-star remnant \citep{Baiotti:2016qnr,Radice:2020ddv}. 
Its initial differential rotation is expected to generate magnetic fields up to $B=10^{16}$\,G, which buoyantly emerge from the remnant and form its external magnetosphere 
\citep{Kluzniak:1997nt,Most:2023sft,Combi:2023yav,Giacomazzo:2013uua,Giacomazzo:2014qba,Kiuchi:2015sga,Mosta:2020hlh,Aguilera-Miret:2023qih}. 
Then, the remnant cools and forms a young magnetar. It likely has a twisted magnetosphere, filled with $e^\pm$ pairs of a small mass density $\rho$. Its expected magnetization parameter $\sigma=B^2/\left(4\pi \rho c^2\right)$ is huge, exceeding $10^{10}$ \citep{Beloborodov:2022pvn}.

Eventually, the remnant emits its angular momentum in a magnetized wind, loses its rotational support, and can collapse into a black hole \citep{Lasky:2013yaa,Ravi:2014gxa,DallOsso:2014hpa}.
The spindown occurs on the timescale $t_{\rm sd}\approx c^3 I/\mu^2\Omega^2\approx  10^4\,\mu_{33}^{-2}(\nu/300\,{\rm Hz})^{-2}$\,s, where $I\approx 10^{45}$\,g\,cm$^2$ is remnant's moment of inertia, $\nu=\Omega/2\pi$ is its rotation rate and $\mu$ is the dipole moment of its magnetosphere (we normalized it to $10^{33}\,$G\,cm$^3$). 
The delayed collapse after $t\sim t_{\rm sd}$ occurs suddenly, 
on a ms timescale.
Proposed electromagnetic signals from the delayed collapse include low-frequency waves \citep{Falcke:2013xpa} and GRBs emitted by collimated jets \citep{Ciolfi:2014yla}.

Previous 
simulations of neutron-star collapse followed the dynamics of the external magnetosphere either in vacuum or using force-free electrodynamics (FFE) \citep{Baumgarte:2002vu,Lehner:2011aa,Dionysopoulou:2012zv,Palenzuela:2012my,Most:2018abt}.
Both frameworks do not allow plasma heating. Furthermore, both neglect plasma inertia and so are unable to track magnetospheric shock formation. By contrast, a full magnetohydrodynamic analysis predicts monster shocks, which 
can generate gamma-rays \citep{Beloborodov:2022pvn}. 

This Letter reports the first magnetohydrodynamic 
simulations
of the magnetosphere evolution in the dynamic spacetime of the collapsing
magnetar.
It demonstrates shocks and ejection of a hot outflow that will emit a GRB. We also find that the outflow carries information about 
quasi-normal modes of the nascent black hole, which may be observed in the GRB time profile.

\section{Methods}
We set up a simple initial state: a dipole magnetic field is attached to a rotating star with $\boldsymbol{\mu}\parallel\boldsymbol{\Omega}$, 
using the vector potential given by \citet{Shibata:2011fj} with a 
surface magnetic field of $B_\ast\sim 10^{16}\, \rm G$.

We artificially reduce the rotation rate from a realistic 
$\nu\gtrsim 300$\,Hz
to $56$ \,Hz, so that the light cylinder $R_{\rm LC}=c/\Omega$ is beyond our computational box. This simplification avoids the challenging preparation of an equilibrium rotating magnetosphere with the equatorial current sheet at $r>R_{\rm RL}$. 
Rotation 
is still essential, as it
enables the ring-down effect when the star collapses into a black hole
(e.g. \citet{Kokkotas:1999bd}).  The amplitude of ring-down oscillations scales with  angular momentum. Therefore, 
we also simulate collapse with a high initial $\nu=900$\,Hz; this additional simulation misses details of the outer magnetospheric dynamics but demonstrates the enhanced ring-down effect on the magnetized outflow from the nascent black hole.

We use the \texttt{RNS} code \citep{Stergioulas:1994ea} to initialize the star
as an unstable general relativistic polytrope (with polytropic coefficient K=100) of mass $M=1.7M_\odot$ and equatorial coordinate radius $R_\star\approx 12$\,km;
the details of its internal structure are unimportant, as we focus on the external magnetosphere. 
The magnetosphere has a low mass density $\rho$ and a large magnetization parameter 
$\sigma\gg 1$.
At time $t=0$, we add a small pressure perturbation and the star begins to collapse.

The simulation tracks the spacetime of the collapsing 
star and evolves its
magnetosphere according to general relativistic magnetohydrodynamics (GRMHD) equations \citep{Duez:2005sf}. Since the plasma magnetosphere has a very small effective resistivity on the dynamical timescales of interests $t_{\rm dyn}\sim R/c$, 
it can be treated
as an ideal conductor everywhere except the sites of magnetic 
reconnection, which 
develops later in the simulation (and is mediated by numerical resistivity).

We use the \texttt{Frankfurt/IllinoisGRMHD (\texttt{FIL})} code \citep{Most:2019kfe,Etienne:2015cea} which is built on top of the \texttt{Einstein Toolkit} infrastructure \citep{Loffler:2011ay}.
The spacetime dynamics is 
tracked using
the Z4c formulation of the Einstein equations \citep{Hilditch:2012fp}. 
We 
use
moving puncture coordinates \citep{Alcubierre:2002kk}
in the simulation and the presentation of results below.
The GRMHD equations  
assume an ideal fluid with thermal pressure proportional to $e-\rho c^2$, where $e$ is the fluid energy density including rest mass. The equations
are evolved using the ECHO scheme \citep{DelZanna:2007pk} with vector potential-based constraint transport \citep{Etienne:2010ui,Etienne:2011re}.

No spatial symmetries are imposed during the simulation.
We employ a fixed three-dimensional Cartesian grid 
with $6$ levels of mesh refinement; the highest resolution has $78$ grid points per $R_\star$.
The grid extends
to $3.25\times R_\star$ ($750\, \rm km$) in each direction.
The boundaries of the computational domain are far from the shock and do not affect the results presented below.

Unlike most MHD simulations in dynamical spacetimes (e.g.
\citealt{Liu:2008xy,Kiuchi:2014hja,Palenzuela:2015dqa,Ciolfi:2017uak,Most:2019kfe}), 
our simulation follows the magnetosphere with a high magnetization parameter (see also \citet{Paschalidis:2014qra,Ruiz:2016rai}).
In particular, the initial ``background'' magnetosphere has $\sigma_{\rm bg}\sim 25$, and its perturbation during the collapse
leads to ultra-relativistic motions with Lorentz factors 
$\gamma\sim 10$.
Such simulations are challenging
 in terms of accuracy requirements and numerical stability of the algorithm,
especially 
when MHD is coupled to
a dynamically evolved spacetime. 

We have 
changed
the code to improve 
its
robustness (see also \citet{Most:2023sme}), including different primitive recovery schemes \citep{Kastaun:2020uxr,Kalinani:2021ofv}, drift floors \citep{Ressler:2016pmh}, and 
bounds on 
$\sigma$ and $\beta= 8\pi P/B^2$
where $P$ is the fluid pressure and 
$B$ is the magnetic field in fluid rest frame.
Specifically, before black hole formation, we enforce bounds, $10< \sigma < 50$, and $\beta > 0.02$, outside of the star. We distinguish between magnetospheric and stellar matter using a passive scalar (see also Ref. \citet{Parfrey:2017nby}).
Even with all these improvements,
we have found 
that the fourth-order derivative corrector 
performs poorly in the shock region. Since the constraint-transport algorithm prevents us from switching it off selectively, we have disabled it everywhere. 
Thus, the simulation maintains second-order accuracy, different from all previous GRMHD simulations carried out with \texttt{FIL} \citep{Most:2019kfe,Most:2021ytn,Chabanov:2022twz,Most:2023sft,Most:2023sme}.

\section{Results}

The evolution observed in the simulation 
may be summarized as follows.

\subsection{Collapse and wave launching}
The collapsing star quickly forms a black hole with the
apparent horizon radius $R_h\approx 2.5\,{\rm km}$ and the ergosphere around it. Effectively, the magnetospheric footprints on the star are quickly pulled in from $R_\star$ to $R_h$, and the star's magnetic flux $\Psi$ is now threading the smaller sphere of radius $R_h$. As a result, a strong quasi-monopolar magnetic field $B\sim \Psi/r^2$ is created in the radial zone $R_h<r<R_\star$.
This inner zone with the amplified magnetic pressure launches a compressive wave into the surrounding magnetosphere, which propagates with nearly speed of light, 
$v_{\rm wave}/c \approx 1-\sigma_{\rm bg}^{-1}$.

\begin{figure}
    \centering
    \includegraphics[width=0.5\textwidth]{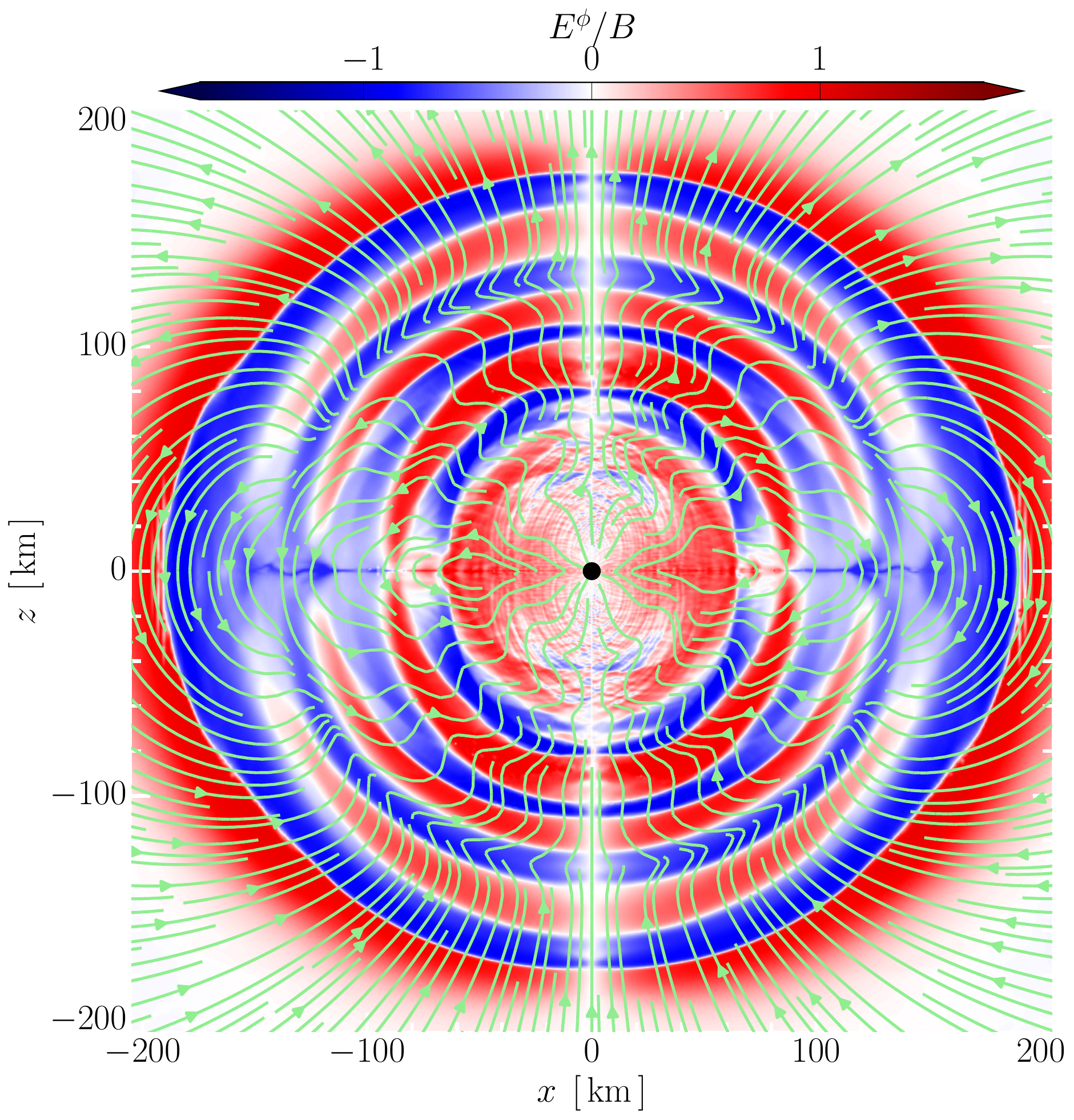}
    \caption{
    Magnetohydrodynamic waves excitepd in the neutron star magnetosphere by the collapse.
    The rotation axis is along $z$, and the figure shows a cut of the magnetosphere along the plane of $y=0$.
    The black circle at the center is the nascent black hole. Green curves show the magnetic field lines, and color shows $E^{\phi}/B$ as measured by the local normal observer. The snapshot is taken at $t=0.67\, \rm ms$ from the onset of collapse.
    }
    \label{fig:Ephi}
\end{figure}

The compressive MHD waves (known as ``fast magnetosonic modes'') have electric field $\bE\parallel \bk\times\bBbg$ where $\bk$ is the (approximately radial) wavevector and $\bBbg$ is the initial background dipole magnetic field. 
So, the wave has a toroidal electric field
$E^\phi$. 
The launched wave of $E^\phi$ continues to propagate outward for the rest of the simulation;
its snapshot 
is shown in Figure~\ref{fig:Ephi} at a late time $t=0.67$\,ms (as measured by a distant observer), near the end of the simulation. The magnetic field ahead of the wave is the dipole $\bBbg$, and the magnetic field behind the wave is close to the split monopole configuration. 
The wave dynamics at radii $r\gg R_h$ can be approximately described neglecting general-relativistic corrections.

\begin{figure*}
    \centering
    \includegraphics[width=\textwidth]{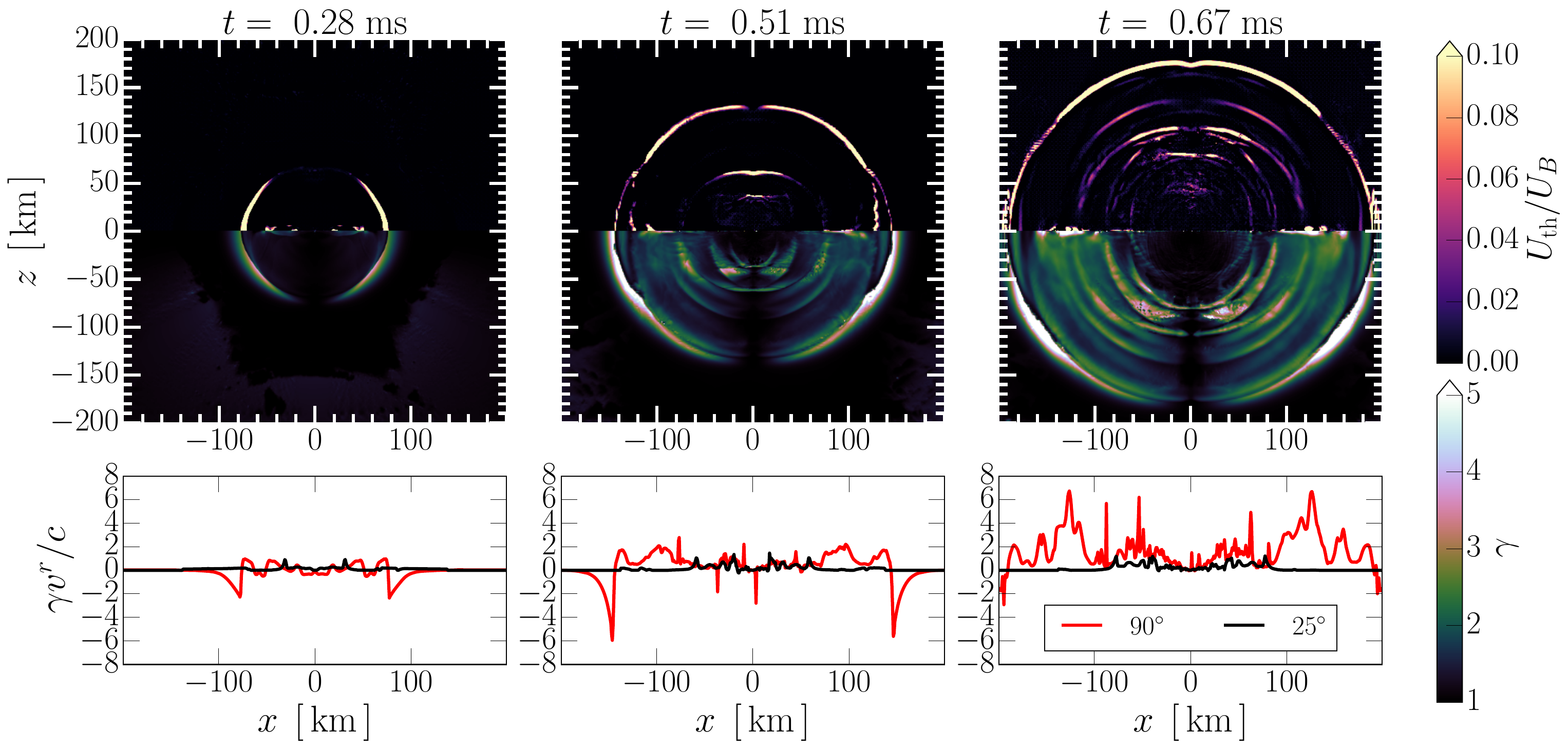}
    \caption{Shock formation and evolution, shown with three snapshots (cuts of the magnetosphere along the vertical plane of $y=0$). Color panels show 
    thermal energy density $U_{\rm th}$ 
    (normalized to the magnetic energy density $U_B$)
    at $z>0$ and fluid Lorentz factor $\gamma$ at $z<0$. Lower panels show the radial four-velocity $u^r=\gamma v^r/c$ measured along two lines in the $x$-$z$ plane: 
    on the $x$-axis (polar angle $\theta=90^\circ$, red) and on the line of $\theta=25^\circ$ (black).
    \\
    \\
    }
    \label{fig:2D_shock}
\end{figure*}

\subsection{Monster shock formation}
Magnetosonic waves were recently shown to accelerate plasma to a huge radial 4-velocity 
$u^r=\gamma v^r/c$ 
\citep{Beloborodov:2022pvn}.
This effect occurs when the wave reaches radius $\Rsh=(c\mu^2/8L)^{1/4}$ in the equatorial plane, where $\mu=r^3B_{\rm bg}$ 
and $L$ is the wave power. 
In the FFE limit ($\sigbg\rightarrow \infty$) fluid expansion in the rarefaction phase of the compressive wave would diverge at $\Rsh$ as $E^2-B^2$ touches zero and $u^r\rightarrow-\infty$. For a finite $\sigbg$, fluid 
develops
a finite $u^r\propto\sigbg$. 
The ultrarelativistic fluid motion is directed toward the star and 
the wave 
immediately
develops a monster shock. 

Shock formation in magnetosonic waves has been demonstrated by kinetic plasma simulations \citep{Chen:2022yci} and by MHD calculations using characteristics \citep{Beloborodov:2022pvn}. For waves with frequency
$\omega\gg c/\Rsh$ a simple analytical MHD solution has been obtained.
It demonstrates that at $r>\Rsh$ the wave profile $E(t-r/c)$ 
develops a plateau of width $W_p\sim c/\omega$
where $E^2\approx B^2$. The plateau
forms a linear accelerator, so the wave 
pushes the fluid 4-velocity to a huge value
\citep{Beloborodov:2022pvn}
\begin{equation}
\label{eq:ur}
   u^r\approx -\frac{ W_p \sigbg}{r}\approx -\frac{c \sigbg}{\omega r}.  
\end{equation}
The accelerated flow dissipates its energy in a monster shock.

The monster shock appears in our simulation where $E^2-B^2$ approaches zero, as predicted. We also
observe the development of an $E$-plateau where $u^r$ develops a steep linear profile, reaching values consistent with Eq.~(\ref{eq:ur}) (see Fig.~\ref{fig:2D_shock}). The shock is the sudden jump of $u^r$ from a large negative value $u^r\sim -6$ back to a moderate $u^r$.
These unique features 
of monster shocks are 
clear in
the simulation despite numerical inaccuracies accumulated in the shock region.
We also observe the expected strong heating localized at the shock (see the temperature panel in Fig.~\ref{fig:2D_shock}). 

The magnetization parameter $\sigbg\sim 25$ used in the simulation is far below its real value in a magnetar, and the shock strength should be scaled to a larger $\sigbg$ according to Eq.~(\ref{eq:ur}). The corresponding large $u^r\propto \sigbg$ will make the shock highly radiative, i.e. the accelerated flow will radiate its energy before crossing the shock and joining the downstream flow \citep{Beloborodov:2022pvn}. Future simulations could attempt to track radiative transfer with self-consistent creation of $e^\pm$ pairs \citep{Beloborodov:2020ylo}, 
which immediately make the flow optically thick. 
The dissipated energy is inevitably thermalized behind the shock, creating 
an
opaque, radiation-dominated outflow. 
Our simulation 
assumes that the released energy remains trapped in the fluid.
The observed
relativistic outflow trails 
the shock, which expands with speed
$c$. 
 \begin{figure*}
    \centering
    \includegraphics[width=\textwidth]{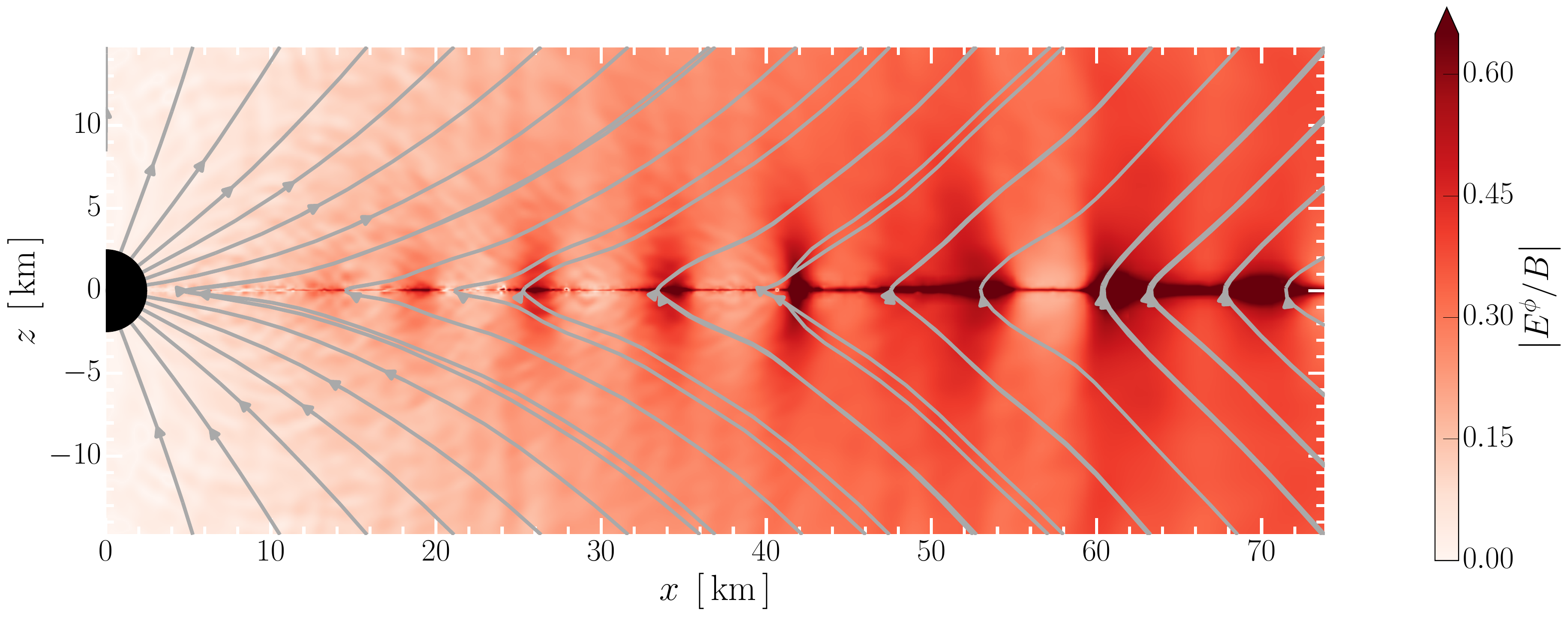}
    \caption{Close-up of the black hole (black circle) and equatorial current sheet with magnetic field lines (gray) 
    after the collapse of a neutron star with initial rotation $\nu=900$\,Hz.  Color shows the azimuthal component $E^\phi$ of the electric field normalized to the magnetic field $B$.
    The snapshot is taken at time $t=0.67\,\rm ms$ 
    after the onset of collapse.
    }
    \label{fig:BH_close}
\end{figure*}
~\\
\subsection{Black hole ring-down}\label{sec:ringdown}
The ring-down of the nascent spinning black hole lasts $\sim 100 R_h/c$. It involves quasi-periodic oscillations of the horizon with frequencies characteristic of black hole quasi-normal modes, whose amplitude decays exponentially with time (e.g.
\citet{Kokkotas:1999bd}). For a stationary black hole, the quasi-monopolar magnetic flux $\Psi$ threading the horizon would be stuck for a significant time --- its decay would  be controlled by magnetic reconnection in the equatorial plane on a timescale $\gtrsim 100R_h/c$ \citep{Bransgrove:2021heo}. By contrast, the oscillating black hole quickly and quasi-periodically sheds magnetic flux, losing significant $\delta\Psi$ each oscillation period. 
The discharged $\delta\Psi$ forms 
a quasi-periodic MHD outflow 
with the characteristic cusps of magnetic field lines in the equatorial plane. 
The cusps are inherited from the earlier split-monopole shape of field lines at the time of their decoupling from the oscillating black hole. 
This effect 
is seen in the simulation with the reduced rotation rate $\nu=56$\,Hz (Fig.~\ref{fig:Ephi}) and becomes stronger in the simulation with fast rotation $\nu=900$\,Hz (Fig.~\ref{fig:BH_close}).

\begin{figure}
    \centering
    \includegraphics[width=0.5\textwidth]{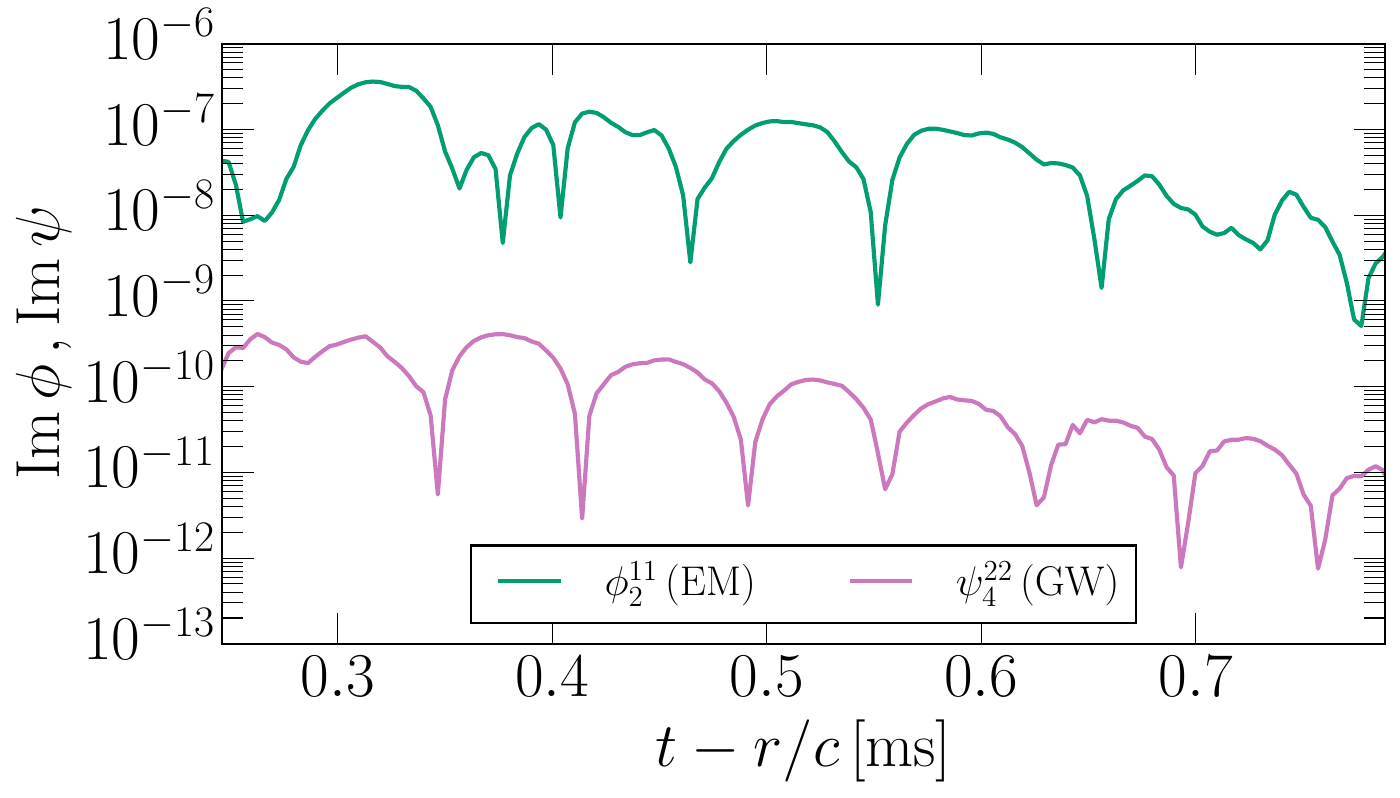}
    \caption{Black hole ring-down 
    of the rapidly rotating configuration, $\nu=900$\,Hz. 
     The ring-down in gravitational waves (GW) and in the electromagnetic field (EM) is shown using the Newman-Penrose scalars $\psi_4$ ($\ell=m=2$ mode) and $\phi_2$ ($\ell=m=1$ mode), respectively. All quantities are measured at a radius $r\approx 75$\,km.
    \\
    } 
    \label{fig:BH_ringdown}
\end{figure}

Previous collapse simulations with a vacuum or FFE magnetosphere showed electromagnetic waves from ring-down  \citep{Baumgarte:2002vu,Lehner:2011aa,Palenzuela:2012my,Most:2018abt}. 
Vacuum ring-down
is usually described as a coupling of black-hole quasi-normal modes to outgoing gravitational and electromagnetic waves \citep{Teukolsky:1972my,Teukolsky:1973ha} using Newman-Penrose scalars
\begin{align*}
     \psi_4 = -C_{\mu\nu\kappa\lambda} n^\mu \bar{m}^\nu n^\kappa \bar{m}^\lambda, \qquad \phi_2 = F^{\mu\nu} \bar{m}_\mu n_\nu.
\end{align*}
Here $C_{\mu\nu\kappa\lambda}$ is the Weyl curvature tensor and $F^{\mu\nu}$ is the electromagnetic tensor; vectors $\bm{m}$ and $\bm{n}$ are conveniently chosen as  $\bm{m} =  (\hat{\bm{\theta}} + i \hat{\bm \phi})/\sqrt{2}$ and $\bm{n} =(\hat{\bm t} - \hat{\bm r})/\sqrt{2}$ (so that $\bm{m}$, $\bm{n}$, and $\bm{l} = (\hat{\bm t} + \hat{\bm r})/\sqrt{2}$ form an orthonormal null tetrad).
Then, $\phi_2\propto E_{{\theta}}+iE_{\phi}$ 
represents two polarization states of the outgoing electromagnetic waves (in MHD, $E_{{\theta}}$ and $E_{{\phi}}$ correspond to the fast magnetosonic and Alfv\'en waves, respectively). 
We have verified that the dominant (quadrupole) component of $\psi_4$ 
observed in our simulation is consistent with the quasi-normal mode computed using \texttt{qnm} code \citep{Stein:2019mop}.
We have also calculated the dominant (dipole) component of $\phi_2$, which also approximately matches the corresponding quasi-normal mode frequency
(Fig.~\ref{fig:BH_ringdown} shows the evolution of dominant spherical harmonics in ${\rm Im}\, \phi_2$ and $\psi_4$).
Note however that $\phi_2$ was designed for vacuum electromagnetic fields, and so the oscillation of $\phi_2$ may not accurately represent the modulation of MHD outflow. The frequency of compressive modulations  observed in Fig.~\ref{fig:BH_close} may be directly estimated as 
$\nu_{\rm mod}=v/\lambda$, where $v\lesssim c$ is the outflow speed and $\lambda\sim 7\,$km
is the spatial modulation period. 
The subsequent evolution of the balding black hole will crucially depend on the resistivity in the current sheet \citep{Bransgrove:2021heo,Selvi:2024lsh}. In the absence of 
a
controlled resistivity in our simulations, we defer 
the study of the late phase
to future work.

\subsection{Gamma-ray burst.}
Our simulation demonstrates that the magnetospheric destruction during the neutron-star collapse involves strong dissipation and creates 
a powerful magnetized
outflow with a characteristic 
peak 
duration $\sim 1$\,ms. The hot outflow is launched behind the leading monster shock and has a quasi-periodic tail. The modulated tail is generated by the nascent black hole, as it rings down and quickly sheds most of its magnetic flux initially inherited from the neutron star.  
The modulation frequency lies naturally in the kilohertz band, suggesting a connection
with recently reported
kilohertz quasi-periodic oscillations in some
GRBs \citep{Chirenti:2023dzl}.

Note that the magnetosphere around the neutron star prior to collapse has a minute plasma density. Therefore, the explosion triggered by collapse 
is practically clean from baryons.
The created $e^\pm$ plasma in the hot outflow is initially opaque to scattering. Most of the heat density $U$ is contained in trapped 
blackbody
radiation $U\approx aT^4$ ($a$ is the radiation constant), as the photons far outnumber the $e^\pm$ pairs.
The outflow expansion to large radii is not followed by our simulation, however its basic features can be predicted in analogy with the well-known ``fireball'' model for cosmological GRBs \citep{Paczynski:1986px,Goodman:1986az}. The outflow Lorentz factor $\gamma$ will grow and its temperature $T$ will drop due to adiabatic cooling.
Eventually, most of $e^\pm$ pairs annihilate, and the trapped photons are released, producing a GRB. The burst duration is set by the outflow duration, which lasts only a few ms after the collapse. 

The burst spectrum will peak at photon energies $\sim 3 k_B T\gamma$, where $k_B$ is the Boltzmann constant. Simplest GRB models assume adiabatic outflows with no magnetic fields, and then $\gamma T\approx const=T_0$ \citep{Paczynski:1986px}. In that case, the observed temperature is weakly changed during the outflow expansion and remains close to the initial temperature. Outflows predicted by our simulation are magnetically dominated, and their temperature may be affected by additional dissipation of magnetic energy at large radii. Regulation of the GRB spectral peak in dissipative outflows is discussed in \cite{Beloborodov:2012ys}.

The energy of the GRB outflow
is set by the pre-collapse energy of the neutron-star magnetosphere $\E_m\sim 10^{47}\mu_{33}^2\,$erg. 
During the collapse
the magnetospheric energy is amplified by compression,
and then most of it is ejected in the outflow. Assuming that $\sim 0.1$ of this energy is eventually emitted in the GRB, we can roughly estimate the GRB luminosity as
$L\sim 0.1\E_m/1\,{\rm ms}\sim 10^{49}\,B_{15}^2\,$erg/s, where $B_{15} = B/10^{15}\, \rm G$ is the magnetic field near the neutron star prior to collapse.
Note that the outflow (and hence the GRB) is anisotropic, but not strongly collimated, 
unlike 
jet-powered GRB models.

During its lifetime prior to collapse, the rotating magnetar produced a magnetic wind
outside the dipole magnetosphere, with a termination shock of a large radius behind the merger ejecta (e.g., \citealt{Metzger:2006mw,Dessart:2008zd})
\footnote{
Energy deposited by the magnetar wind into the merger ejecta was proposed to power an optical transient much brighter than a usual kilonova \citep{Metzger:2013cha}, and non-detection of this effect can constrain the magnetar lifetime \citep{Wang:2023qww}.
}.
The collapse launches a relativistic shock, which will continue to expand into the cold wind. 
The shock
is expected to emit a fast radio burst (FRB) at radii $r\sim 10^{13}$\,cm 
\citep{Beloborodov2020_FRB}.
This suggests a mechanism for a delayed FRB from neutron star mergers, emitted together with the delayed GRB. However, the FRB can hardly escape through the surrounding shell of mass $M_{\rm ej}\sim 10^{-2}M_\odot$ ejected earlier by the merger \citep{Bhardwaj:2023avo,Radice:2023zfv}. A similar problem is faced by the recently proposed FRB-GRB connection \citep{2023arXiv231204237R}.

The 
massive shell is 
also a 
threat
to the GRB predicted by our simulation. The
GRB 
will not be
blocked by $M_{\rm ej}$ if the 
magnetar collapse occurs
with a sufficient delay $\Delta t\gtrsim 10$\,hr after the merger, so that the massive ejecta expands to radius $R\sim 3\times 10^{14}\,(\Delta t/10{\rm\,hr})$\,cm and its optical depth to gamma-rays $\tau\sim \kappa M_{\rm ej}/4\pi R^2$ drops below unity ($\kappa$ is the ejecta opacity to Compton scattering).

\section{Acknowledgments}
The authors gratefully acknowledge insightful discussions with A. Levinson, M. Lyutikov, A. Philippov, V. Ravi, R. Sari, L. Sironi, J. Stone, S. Teukolsky and C. Thompson. 
AMB and ERM acknowledge support from NASA's ATP program under grant 80NSSC24K1229.
ERM acknowledges support by the National Science Foundation under grants No. PHY-2309210 and AST-2307394.  
A.M.B. is supported by NSF AST~2009453, NASA~21-ATP21-0056 and Simons Foundation award No. 446228.
B.R. is supported by the Natural Sciences \& Engineering Research Council of Canada (NSERC) and by a grant from the Simons Foundation (MP-SCMPS-00001470). 
Simulations were performed on the NSF Frontera supercomputer under grant
AST21006, and on Delta at the National Center for
Supercomputing Applications (NCSA) through allocation PHY210074 from the
Advanced Cyberinfrastructure Coordination Ecosystem: Services \& Support
(ACCESS) program, which is supported by National Science Foundation grants
\#2138259, \#2138286, \#2138307, \#2137603, and \#2138296.

\software{EinsteinToolkit \citep{Loffler:2011ay},
      Frankfurt/IllinoisGRMHD \citep{Most:2019kfe,Etienne:2015cea}
      RNS \citep{Stergioulas:1994ea},
      kuibit \citep{kuibit},
	  matplotlib \citep{Hunter:2007},
	  numpy \citep{harris2020array},
      qnm \citep{Stein:2019mop},
	  scipy \citep{2020SciPy-NMeth}
}

\bibliography{inspire,non_inspire}%

\end{document}